\documentclass[manuscript]{aastex}

\slugcomment{Revised for The Astronomical Journal}

\shorttitle{Haumea's Moons}
\shortauthors{\' Cuk, Ragozzine \& Nesvorn\' y }

\begin{document}

\title{On the Dynamics and Origin of Haumea's Moons}

\author{Matija \' Cuk}
\affil{Carl Sagan Center, SETI Institute, \\
 189 North Bernardo Avenue, Mountain View, CA 94043\\
\email{mcuk@seti.org}}

\author{Darin Ragozzine}
\affil{Department of Astronomy, University of Florida, \\ 
Gainesville, FL 32611}

\and

\author{David Nesvorn\' y}
\affil{Southwest Research Institute, \\
Boulder, CO 80302}

\begin{abstract}
The dwarf planet Haumea has two large satellites, Namaka and Hi'iaka, which orbit at relatively large separations. Both moons have significant eccentricities and inclinations, in a pattern that is consistent with a past orbital resonance \citep{rag09}. Based on our analysis, we find that the present system is not consistent with satellite formation close to the primary and tidal evolution though mean-motion resonances. We propose that Namaka experienced only limited tidal evolution, leading to the mutual 8:3 mean-motion resonance which redistributed eccentricities and inclinations between the moons. This scenario requires that the original orbit of Hi'iaka was mildly eccentric; we propose that this eccentricity was either primordial or acquired though encounters with other TNOs. Both dynamical stability and our preferred tidal evolution model imply that the moons' masses are only about one half of previously estimated values, suggesting high albedos and low densities. As the present orbits of the moons strongly suggest formation from a flat disk close to their present locations, we conclude that Hi'iaka and Namaka may be second-generation moons, formed after the breakup of a past large moon, previously proposed as the parent body of the Haumea family \citep{sch09}. We derive plausible parameters of that moon, consistent with the current models of Haumea's formation \citep{lei10}. An interesting implication of this hypothesis is that Hi'iaka and Namaka may orbit retrograde with respect to Haumea's spin. Retrograde orbits of Haumea's moons would be in full agreement with available observations and our dynamical analysis, and could provide a unique confirmation the "disrupted satellite" scenario for the origin of the family. 
\end{abstract}

\keywords{celestial mechanics -- Kuiper belt -- minor planets, asteroids -- planets and satellites: formation}

\section{Introduction}

Haumea is one of the largest trans-Neptunian objects \citep[TNOs;][]{bro05, rab06}. Together with Pluto, it is the only TNO known to have multiple satellites, with two known moons Hi'iaka and Namaka \citep{bro06}, compared to Pluto's five \citep{sho12}. Unlike somewhat larger Pluto, Haumea is known to be a significantly elongated rapid rotator \citep{rab06}, which is unusual among large TNOs. Another singular feature of Haumea is its dynamical family \citep{bro07, rag07}, the only known compositional cluster in the Kuiper Belt. The family is well established due to conspicuous water-ice-dominated surface composition of Haumea and its family members \citep{rab06, bar06, bro07}. Unlike the main-belt asteroid families \citep{nes06, mar11}, the velocity dispersion of Haumea family members is significantly smaller than the escape velocity of Haumea, requiring an explanation \citep{lyk12, vol12}. 

These numerous puzzling characteristics of Haumea, its satellites and its family have challenged theorists and led to several hypotheses on their origin. \citet{lev08} proposed an origin of Haumea's family in a collision between two large scattered disk objects, an event which  should happen about once during the age of the Solar System. However, a high-velocity collision as proposed by \citet{lev08} does not explain most of the peculiarities of Haumea and the family, including the low velocity dispersion. \citet{sch09} proposed that the family was not formed by an impact on Haumea, but on its past large satellite, and that both the existing moons and family members of Haumea represent fragments of this "ur-satellite". The smaller size of the satellite (compared to Haumea) would explain the low velocity dispersion of the Haumea family, and the present large orbital radii of Namaka and Hi'iaka reflect the past large-scale tidal evolution of the much larger ur-satellite. However, a large ur-satellite could only tidally evolve by removing angular momentum from Haumea's rotation, which needs to be reconciled with its current rapid spin (see Section 6).

\citet{lei10} proposed that Haumea and its family were formed in a slow collision of two large bodies, and showed that such a collision would naturally result in a rapidly-spinning elongated body and a low-velocity-dispersion family, with the bound debris eventually forming the satellites. The most serious argument against the \citet{lei10} scenario is the low probability of a low-velocity collision in the excited Kuiper Belt. As the dynamical family appears to be largely intact, it is impossible for the family-forming collision to have happened before the planetary migration and final sculpting of the Kuiper Belt.\footnote{An interesting possibility that needs to be studied further is that the slow collision happened between two components of a perturbed TNO binary \citep{mar11}.} \citet{ort12} proposed a more general model that involves rotational breakup of proto-Haumea, with the angular momentum added through collisions.

With the origin of Haumea's shape, spin, and family still being debated, the orbits of its two satellites could be used to constrain the past evolution of the system. Table \ref{elem} lists "nominal" orbital and physical parameters of the system that are mostly based on \citet[][ hereafter RB09]{rag09}, with some updates (Ragozzine \& Brown, in preparation). The two satellites are relatively large, with Hi'iaka having a nominal mass that is 0.5\% of Haumea's mass ($M_H$), and Namaka being about an order of magnitude less massive. Both Namaka and H'iaka orbit at significant distance from Haumea, with semimajor axes of 36 and 69 Haumea radii ($R_H$), respectively\footnote{Haumea is significantly tri-axial. Unless otherwise stated, we will use $R_H$ for the mean radius (718~km).}. Namaka has significant eccentricity and inclination ($0.2$ and $13^{\circ}$) while Hi'iaka's orbit is less excited ($e=0.05$ and $i=1-2^{\circ}$). The inverse correlation between the eccentricities and inclinations and the moons' masses is consistent with the orbital excitation through (possibly resonant) interaction between the satellites. In particular, \citet{rag09} suggested that the moons may have passed through their mutual 3:1 mean motion resonance in the past, which could be the source of eccentricities and inclinations. Presently, the two bodies are either close or in their 8:3 mean-motion resonance. 

\defcitealias{rag09}{RB09}


\begin{table}
\begin{center}
\caption{Orbital elements of Haumea's moons. Values taken from \citetalias{rag09}, unless marked with asterisk, in which case more recent data were used (Ragozzine \& Brown, in preparation). Masses listed here (in units of Haumea's mass) are referred in the text as "nominal", inclinations are measured with respect to Haumea's equator, and the semimajor axes are expressed in terms of Haumea's mean radius. Errorbars on masses, semimajor axes and Hi'iaka's eccentricity are from \citetalias{rag09}. Note that the "nominal" masses are in reality much less well constrained than the formal errorbars suggest (Ragozzine \& Brown, in preparation).}
\bigskip
\begin{tabular}{crrrr}
\tableline\tableline
Moon & Mass ($M_H$) & Semimajor axis ($R_H$) & Eccentricity & Inclination ($^{\circ}$)\\
\tableline
Namaka &$5 \pm 4 \times 10^{-4}$ & $35.7 \pm 0.1 $  & 0.2 $^*$ & $13 \pm 2^*$\\
Hi'iaka & $5 \pm 0.3 \times 10^{-3}$ & $69.5 \pm 0.3$ & $0.051 \pm 0.008$& $\simeq 2^*$\\
\tableline
\end{tabular}
\end{center}
\label{elem}
\end{table}

In the following four Sections, we will discuss three distinct hypotheses about the origin of Haumea's moons' eccentricities and inclinations. First, we will assume that they formed on circular planar orbits, with Namaka starting interior to Hi'iaka's 3:1 mean-motion resonance (MMR) and subsequently tidally evolving into the resonance (Section 2). Second, we will assume formation on "cold" (circular and planar) orbits between 3:1 and 8:3 resonances, followed by slow tidal evolution into the 8:3 resonance (where the moons may still reside; Sections 3 and 4). Third, we will consider minimal tidal evolution of the moons on already excited ("hot") orbits into the 8:3 resonance (Section 5). We find that the last hypothesis fits the constraints best, and in Section 6 we will discuss implications of that orbital history for the origin of Haumea, its moons and its dynamical family.

\section{Past 3:1 Mean-Motion Resonance?}

The possibility of past (or current) resonances between the satellites, first suggested by \citetalias{rag09}, is attractive for several reasons. Regular satellites are thought to form on circular and planar orbits, and resonances are a natural way of producing eccentricities and inclinations in a multiple-satellite system \citep{can99}. The smaller moon Namaka has larger eccentricity and inclination than the larger one, Hi'iaka, and the eccentricities and inclinations of each moon appear to be comparable: $e=0.2$ and $\sin i =0.22$ for Namaka; with $e=0.05$ and $\sin i = 0.02-0.04$ for Hi'iaka. If we calculate angular momentum deficit \citep[AMD; ][]{las97} 
\begin{equation}
\Delta=m \sqrt{a}(1-\sqrt{1-e^2} \cos{i}) \simeq m \sqrt{a} {1/2} (e^2 + \sin^2{i})
\end{equation}
for each body, it appears that AMD is approximately equally distributed between the two bodies and between their eccentricities and inclinations. This implies strong dynamical coupling between the two bodies, most likely through an orbital resonance. The fact that both eccentricity and inclination are affected by this interaction points to a resonance of at least second order, as first-order resonances do not affect inclinations \citep{md99}. 

\citetalias{rag09} proposed that the two satellites acquired their eccentricities and inclinations though past capture into the mutual 3:1 MMR. This hypothesis is in agreement with the expected convergent orbital evolution of the two moons due to tides \citep{md99, can99}. The orbits are expected to converge as the rate of tidal evolution depends on the distance as $a^{-5.5}$ and only linearly on mass. Namaka, which is less massive by a factor of ten but is twice closer to Haumea, is expected to migrate outward several times faster than Hi'iaka. Also, the 3:1 MMR is of the second order and thus can affect both eccentricities and inclinations, as required by the present orbits.

\begin{figure}
\epsscale{.75}
\plotone{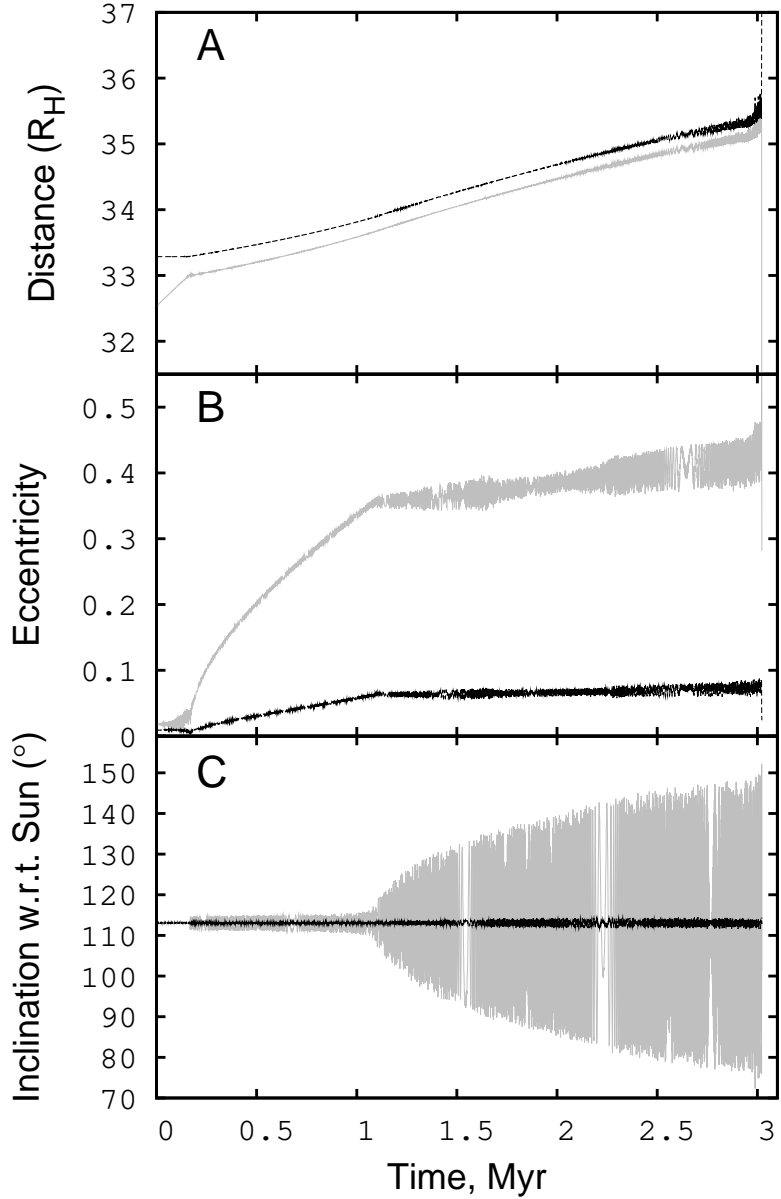}
\caption{Evolution of Namaka (gray) and Hi'iaka (black) through their mutual 3:1 resonance, using nominal masses. Panel A plots semimajor axis of Namaka and the location of exact 3:1 mean-motion commensurability with Hi'iaka (which differs slightly from physically relevant resonant arguments), while panels B and C plot the moons' eccentricity and inclination relative to Haumea's heliocentric orbit. Namaka's tidal evolution in this simulation is orders of magnitude faster than a physically plausible rate. Periodic "gaps" in the plots (especially for Namaka's inclination) are artifacts of the output frequency and filtering.}
\label{31full}
\end{figure}

To test this hypothesis, we performed numerical simulations of the moons' tidal evolution. Throughout this paper we used a custom-made numerical integrator developed by \citet{cuk09}. The integrator uses symplectic principles pioneered by \citet{wis91}, specifically the "democratic" implementation proposed by \citet{lev94}. Given the dominance of Namaka's tidal evolution, we ignored tidal recession of Hi'iaka. We also ignored satellite tides, which are unlikely to significantly affect the eccentricity during limited tidal evolution dominated by resonant excitation. Unless stated otherwise, we used $J_2=0.25$ for the second zonal coefficient of oblate Haumea.

Figure \ref{31full} shows an example of evolution through the 3:1 resonance. Namaka has been initially placed at the semimajor axis lower than that of the resonance, and its rate of tidal evolution has been greatly exaggerated to speed up the evolution (other simulations with slower evolution rates produced the same result). The two moons are at first captured into a mutual eccentricity-type 3:1 mean-motion resonance. In this resonance, eccentricities of both moons increase while the semimajor axis ratio stays constant. Halfway through the resonant evolution, higher eccentricities lead to resonance overlap, meaning that the system is in multiple sub-resonances (involving both eccentricity and inclination) of the 3:1 resonance. Resonance overlap leads to increase in mutual inclination until the point where the resonance becomes unstable due to high eccentricities and inclinations. While resonances typically prevent close approaches between the two bodies, even if their orbits are crossing \citep{md99}, large librations accompanying high-eccentricity resonant configurations can lead to close approaches and scattering \citep{can99}. It is clear that this mechanism of resonance breaking cannot lead to the present, stable system. Unsurprisingly, the eccentricities at which instability happens are significantly higher than observed, even if the pattern of relative eccentricities and inclinations of the two moons roughly matches to the observed state.

\begin{figure}
\epsscale{.75}
\plotone{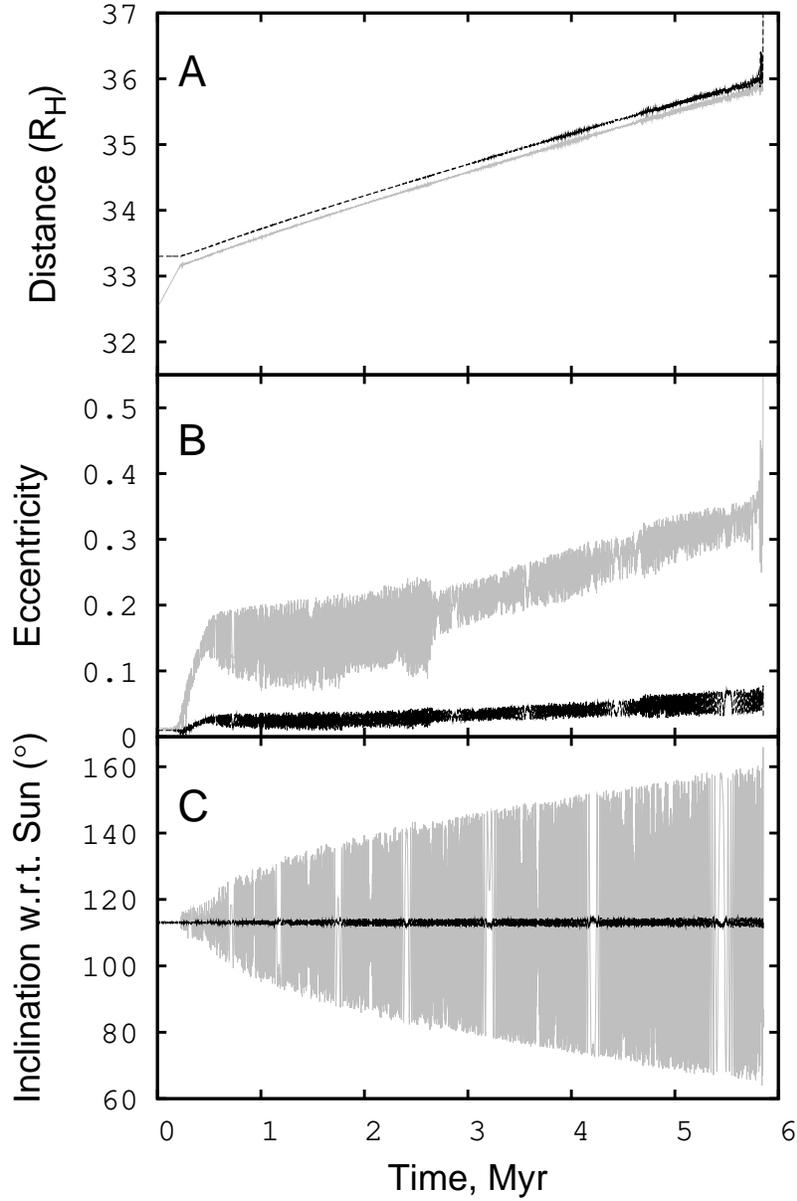}
\caption{Evolution of Namaka (gray) and Hi'iaka (black) through their mutual 3:1 resonance, using one half of their nominal masses. Quantities plotted and simulation parameters were otherwise the same as in Fig. \ref{31full}.}
\label{31half}
\end{figure}

Since \citetalias{rag09}, new work by the same authors indicates that the satellite masses may not have been firmly detected in the original analysis (Ragozzine \& Brown, in preparation). As the masses of the two moons are currently uncertain, we also integrated the system through the 3:1 resonance using the same mass ratio but lower masses for the moons (in line with the expectation that Hi'iaka and Namaka should have similar albedos and densities). Figure \ref{31half} shows the evolution of the moons through the resonance assuming each had one half of the nominal mass estimated by \citetalias{rag09}. The instability happens nonetheless and there appears to be no way of reaching the present system by evolution through the 3:1 resonance. We found that this holds for all masses between 40\% and 100\% of the \citetalias{rag09} nominal mass. Masses smaller than 40\% of nominal masses are not impossible, but start to require extreme albedos and densities even compared to the expected high albedos of these moons  \citepalias{rag09}. Such low masses also imply very weak tidal evolution (see below). 

There is another, purely analytical argument against past evolution through the 3:1 resonance. As \citetalias{rag09} pointed out, large-scale tidal evolution of Haumea's moons (especially Hi'iaka) over the age of the Solar System would require surprisingly strong tidal response of Haumea. Even if Haumea deforms in response to tides as much as a fluid body (i.e. has a Love number of $k_2=1.5$), tidal quality factor would have to be $Q \simeq 10$ \citep{md99}. A more realistic Love number of $k_2=0.1$ would imply $Q \simeq 1$, which is unphysical (as $Q$ is the {\it inverse} of the fraction of deformation energy lost during a cycle). These considerations depend sensitively on the physical size of Haumea (as the rate of tidal evolution is inversely proportional fifth power of radius), and can be changed substantially if Haumea is found to be significantly larger and less dense than currently thought.

To have any substantial orbital evolution of Namaka alone, $Q < 10$ is still required for Haumea in order for the moons first to evolve through the 3:1 resonance and then for Namaka to subsequently migrate to its present location. A more realistic (if optimistic) $Q=50$ allows for $\Delta a/a \simeq 0.02$, which is comparable to the distance between the 3:1 resonance and its current orbit. Using an equatorial rather than mean radius for tidal calculations increases relative semimajor axis evolution of Namaka to $\Delta a/a \simeq 0.05$ (assuming an age of 4.5 Gyr and $Q=50$ for Haumea), which may be marginally consistent with past capture into 3:1 resonance (requiring $\Delta a/a \simeq 0.07$), but then we are again confronted by a lack of a plausible mechanism to disrupt the resonance. 

Our analysis here assumes that Hi'iaka and Namaka were the only two satellites of Haumea present at the time of the 3:1 resonance. Given the large-scale excitation of Namaka's orbit, if an additional moon was present it could have easily been destabilized, breaking the resonance in the process. This hypothesis could in principle address the origin of the orbital excitation in the system, but the unconstrained parameters of the putative third satellite make it hard to test. For reasons of simplicity, we currently prefer the two-body resonant interactions to scenarios involving extinct moons as the explanation for the observed orbital excitation. Ultimately, even if 3:1 resonance crossing can be shown to be plausible (with or without a third satellite), the greater issue of the impossibility of Hi'iaka's orbital evolution to 69 $R_H$ remains unresolved. This seriously undermines the hypothesis that Namaka and Hi'iaka formed in a compact disk following a giant impact.

To summarize, a past 3:1 resonance cannot explain present system but always leads to instability, and may be unlikely from the point of view of tidal physics. A simple estimate of past tidal evolution of Namaka is consistent with a formation between 3:1 and 8:3 resonances. Therefore, we will not discuss the 3:1 resonance any further and will turn to attention to the 8:3 MMR.

\section{Current 8:3 Mean-Motion Resonance?}

\citetalias{rag09} mention that their orbit solutions allow the moons to be in their mutual 8:3 resonance, or at least chaotically enter the resonance at times. Here we do not try to fit the actual observations, but focus on the plausibility of a long-lived 8:3 resonance. While small differences in initial conditions can place the moons inside or outside 8:3 resonance, the resonant and non-resonant long-term evolutions diverge drastically. 

\begin{figure}
\epsscale{.75}
\plotone{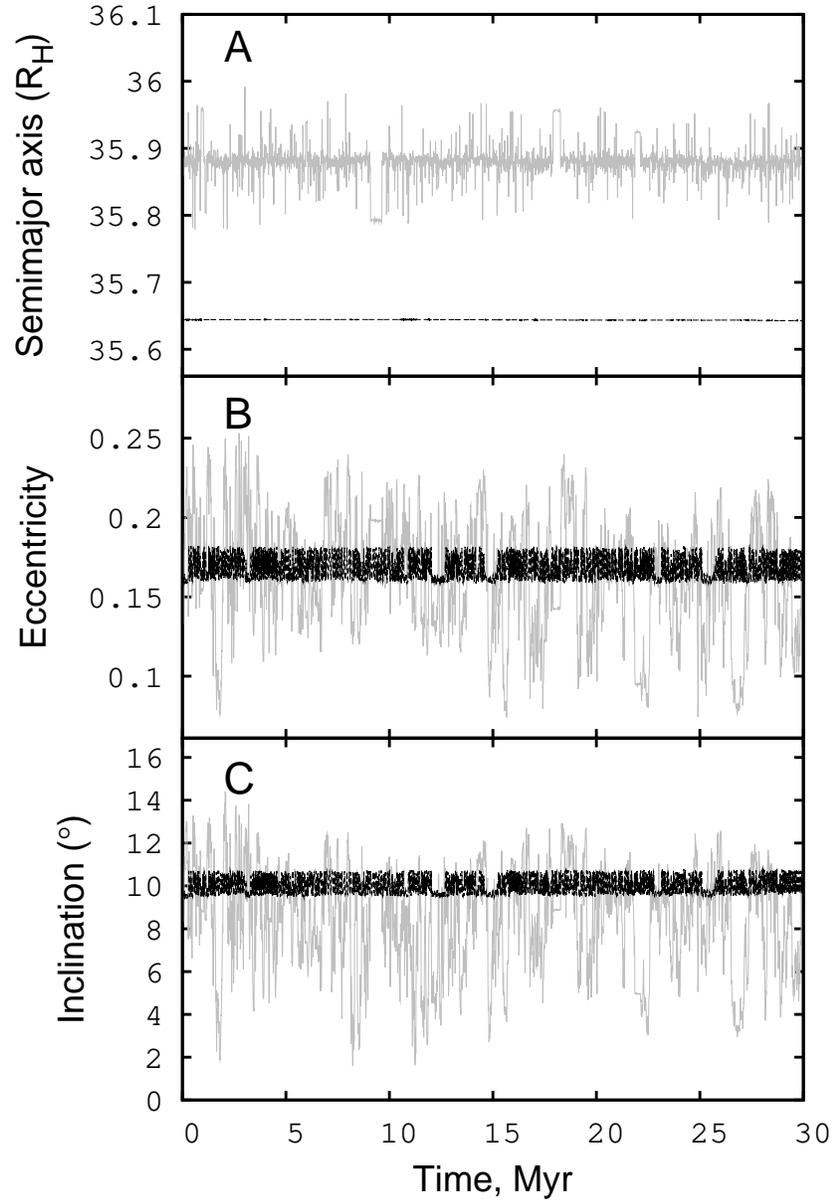}
\caption{A 30-Myr simulation of Namaka's orbit using initial conditions placing the orbit inside (gray) and outside (black) the moons' 8:3 mean-motion resonance. The panels plot Namaka's (averaged) semimajor axis (A), eccentricity (B) and inclination with respect to Haumea's equator (C). Both sets of initial conditions are within uncertainties reported by \citetalias{rag09}. }
\label{namaka}
\end{figure}

\begin{figure}
\epsscale{.75}
\plotone{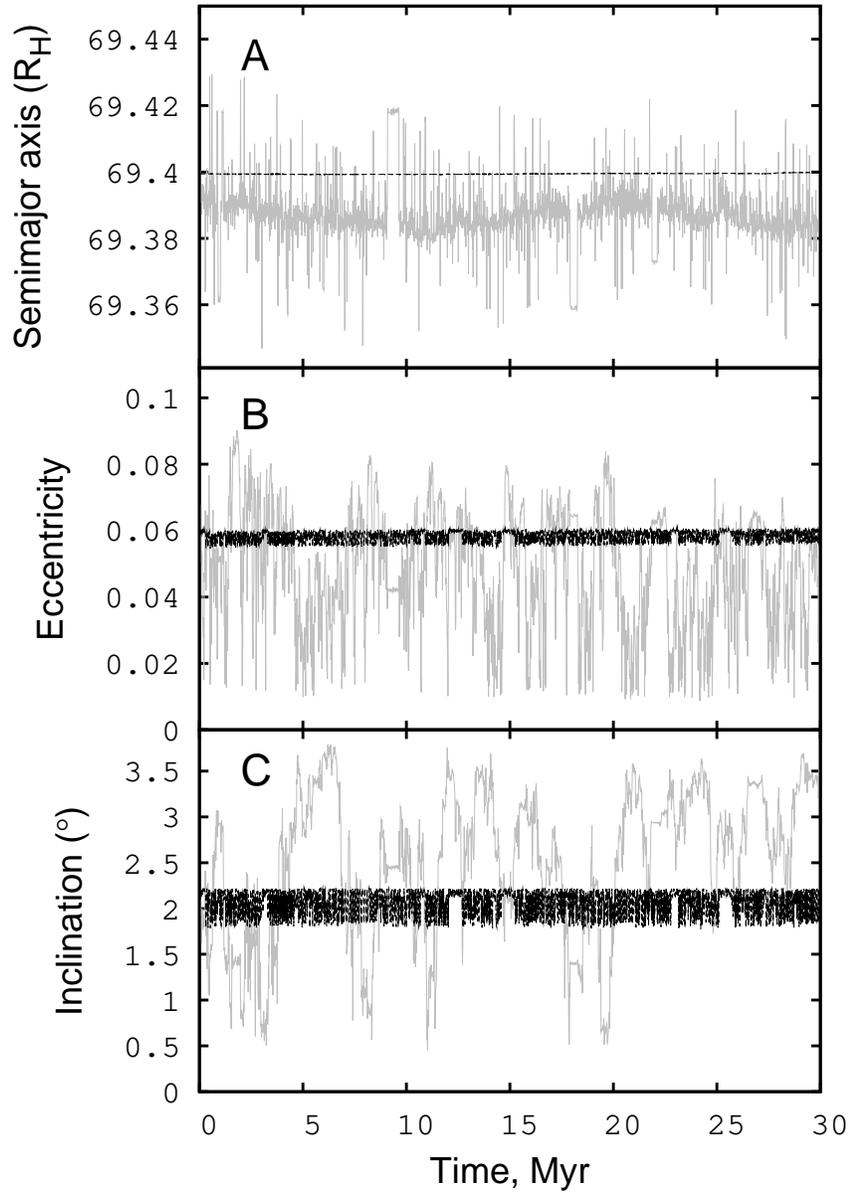}
\caption{Same as Fig. \ref{namaka}, but for Hi'iaka. The panels plot Hi'iaka's semimajor axis (A), eccentricity (B) and inclination with respect to Haumea's equator (C), for one 8:3 resonant (gray) and one non-resonant orbit (black) consistent with observations.}
\label{hiiaka}
\end{figure}

\begin{figure}
\epsscale{.6}
\plotone{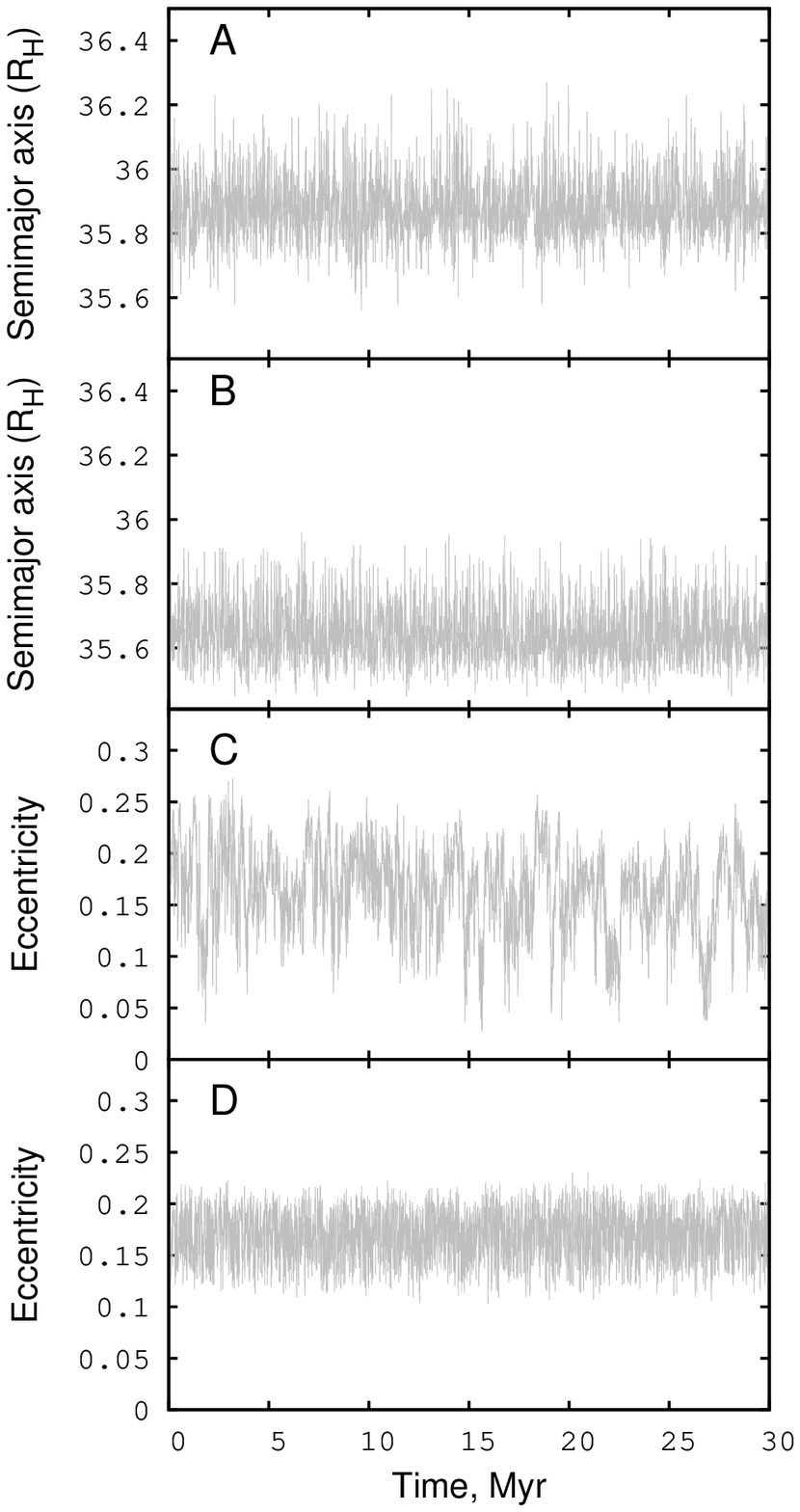}
\caption{Osculating (i.e. non-averaged) orbital elements for Namaka in the resonant and non-resonant solutions (cf. Fig \ref{namaka}). Panels A and B plot the semimajor axis and panels C and D eccentricities of the resonant (A, C) and non-resonant (B, D) solutions, respectively. In contrast to Fig \ref{namaka}, where averaged semimajor axis suggests that the solutions are well-separated, osculating elements clearly overlap, with the resonant orbit having a larger range of variation of orbital elements than the non-resonant one.}
\label{namaka_nofil}
\end{figure}

Figures \ref{namaka} and \ref{hiiaka} show the evolution of semimajor axis, eccentricity and inclination of Namaka and Hi'iaka for two orbital configurations that are within the stated uncertainties from the solution of \citetalias{rag09}. While at first glance the solutions appear very distant from each other, Figure \ref{namaka_nofil} illustrates the osculating (non-averaged) elements of these two solutions do overlap. In these simulations we assumed nominal masses for the satellites and Haumea's oblateness of $J_2=0.25$ (calculated using the mean radius). It is clear that the orbital elements vary more widely when the two orbits are within the 8:3 resonance, than when they are just interior to the resonance. The chaos in the resonant solution is clearly caused by the overlap of many sub-resonances of the 8:3 resonance \citep{md99}. Large eccentricities and inclinations of both bodies put the system in many overlapping sub-resonances at the same time. Taking into account D'Alembert rules \citep{md99}, we find that the 8:3 resonance contains 28 separate sub-resonances, the overlap of which leads to the observed chaos.

While we label the more stable solution "non-resonant", it actually does show modest effects of an outlaying secondary resonance on $e$ and $i$ (seen as changes in the range of $e$ and $i$ variation in Figs. \ref{namaka} and \ref{hiiaka}). This solution remains stable over tens of Myr and there is no reason to doubt its stability on the age of the Solar System.  The resonant solution is much harder to characterize and long-term integrations are required to determine if it is stable. Using orbital elements from \citetalias{rag09} (with Namaka's eccentricity updated to $e_1=0.2$), we ran a number of numerical integrations of the system. In these long-term integrations we varied the mass of the satellites as the most uncertain parameter (while we assumed fixed Hi'iaka-Namaka mass ratio) and for each satellite mass we also ran a number of integrations with slightly different initial conditions. As we used a pair of multi-processor workstations, billion-year integrations would take longer than the time available for this project. Instead, we employed multiple $<3 \times 10^8$~yr integrations with slightly different initial conditions as a probe of longer-term behavior. If we assume that the dynamics of these "clone" integrations becomes uncorrelated after several Myr, we reason that stability in, for example, ten integrations of 100 Myr is roughly equivalent to demonstrating the stability over 1 Gyr. While these are reasonable assumptions, some of the resonant perturbations may be cumulative and multi-Gyr integrations are still required to definitely confirm the lifetimes estimated using multiple shorter runs. Usually we ran a set of simulations for a particular satellite mass only until an instability was observed, after which we re-purposed the processors to integrations using satellite masses that were still consistent with stability. Note that any simulations of the moons' dynamics longer than 100 Myr should also include tidal or passing-body effects (see Section 5).

\begin{figure}
\epsscale{.75}
\plotone{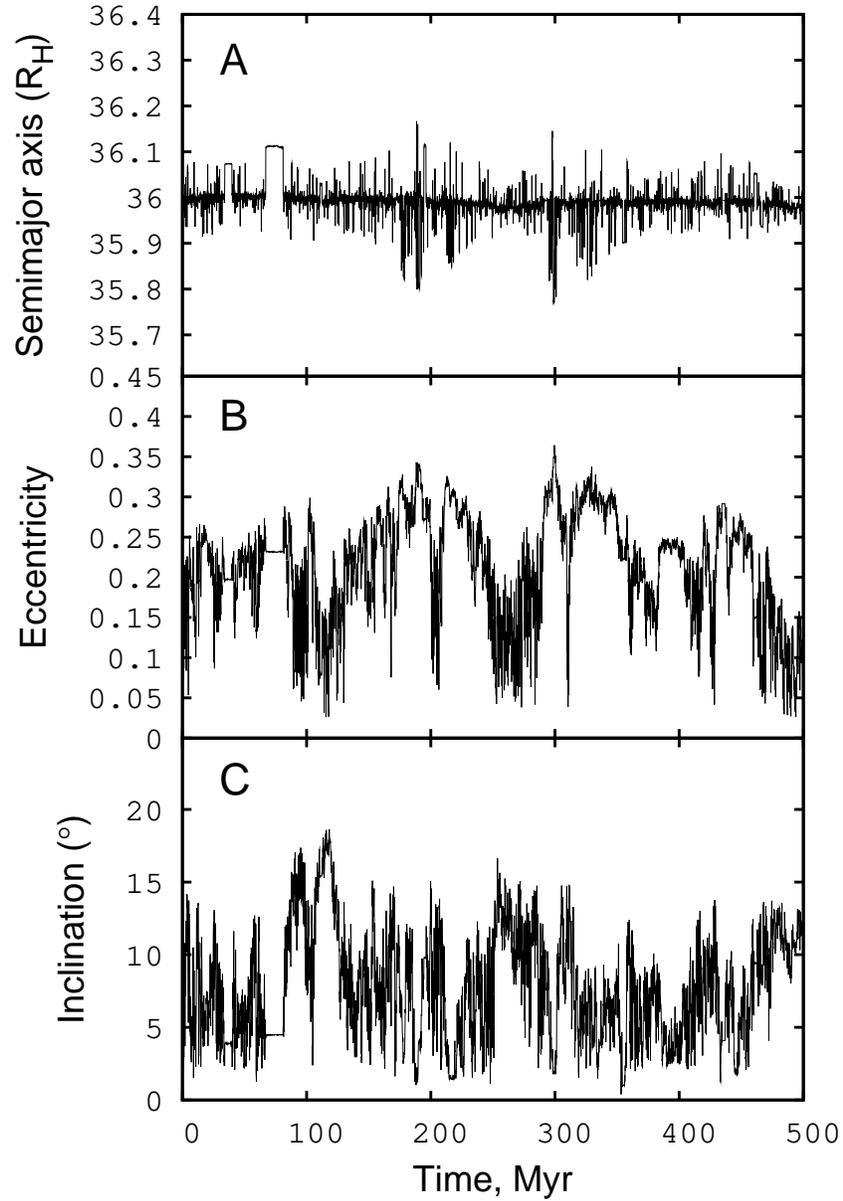}
\caption{A 500-Myr integration of a solution putting Namaka and Hi'iaka in an 8:3 resonance, assuming masses that are 50\% of the nominal ones. Panels A, B and C show Namaka's semimajor axis, eccentricity and inclination, respectively. We ran seven similar simulations using slightly different initial conditions, with a total run-time of 3 Gyr using these masses, and in all cases the system remained stable.}
\label{longterm5}
\end{figure}

\begin{figure}
\epsscale{.75}
\plotone{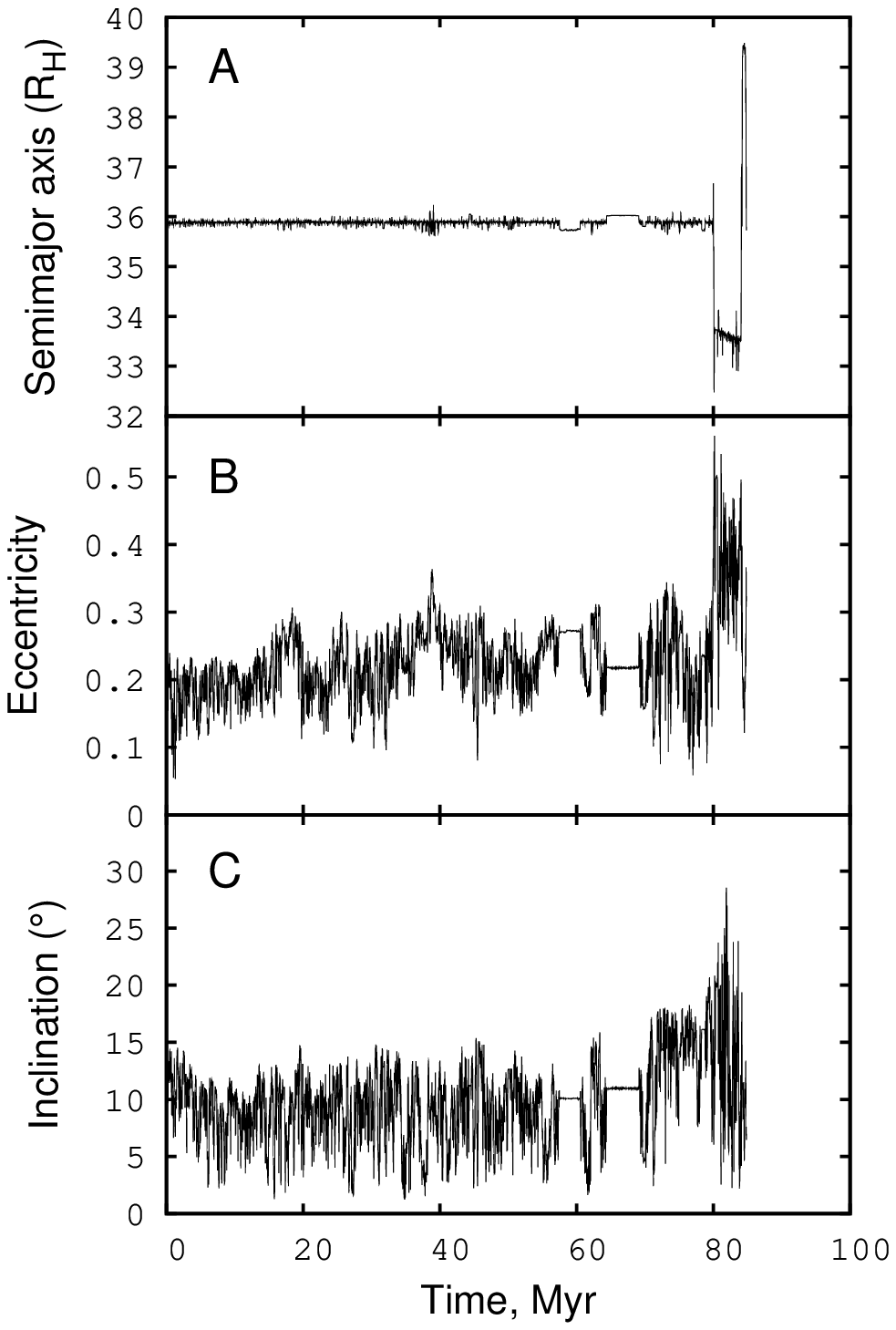}
\caption{A simulation of a Namaka-Hi'iaka 8:3 resonance assuming nominal masses from \citetalias{rag09}. Panels A, B and C show Namaka's semimajor axis, eccentricity and inclination, respectively.The satellites become unstable after about 85 Myr. In numerous simulations we find that the system always goes unstable in about 100 Myr if the satellites have masses that are 70\% or more of those in the nominal solution.}
\label{unstable}
\end{figure}

We find that the masses of Hi'iaka and Namaka that are about 50\% of nominal allow for stability beyond a Gyr (Fig. \ref{longterm5}). Satellite masses set at 60\% of the nominal value lead to instability after about 1 Gyr of combined integrations, and higher masses consistently cause instability after about 100 Myr (Fig. \ref{unstable}).  While the satellite masses that are 60\% of nominal cannot be excluded (especially since it is unknown when the resonance may have been established), we prefer the estimate of satellite masses at 50\% of the nominal solution of \citetalias{rag09}. Even if the moons are not currently in resonance, their masses still need to be in this low range if they ever spent a significant amount of time ($>100$ Myr) in the resonance and survived. 

This suggestion that satellite masses are much lower than nominal is actually not surprising, as unpublished new data (Ragozzine \& Brown, in preparation) suggest that the original analysis overestimated the confidence with which the satellite masses were detected in the fit. About 50\% of the nominal mass is on the low end of the range expected for real Solar System bodies, and implies very high albedos ($0.9$ or so) and very low densities ($\rho= 0.5~\rm{g~cm^{-3}}$) for the moons. The above numbers suggest a relatively pure water-ice composition and a highly porous internal structure; this is consistent with surface composition \citep{sch08} and albedos \citep{ell10} of observed Haumea family members. Additionally, similar low densities have been determined for binary TNOs unrelated to Haumea \citep{nol08}. As the low densities we propose for Haumea's moons likely apply to all Haumea family members, this has important implications for the total mass in the Haumea family.

\section{Large-Scale Tidal Evolution within the 8:3 Mean-Motion Resonance?}

Numerical simulations suggest that the moons may presently be in the 8:3 MMR, and this resonance is stable if the moons' densities are about 0.5~$\rm{g~cm^{-3}}$. This resonance could explain the pattern of their eccentricities and inclinations, but could it also be the sole source of the moons' angular momentum deficit? We ran a large number of tidal evolution simulations which evolved two low-eccentricity ($e\lesssim0.01$) coplanar orbits just wide of the 8:3 MMR into the resonance. Unlike the runs involving capture into the 3:1 resonance in the previous section, the low strength of the 8:3 resonance necessitated much slower tidal evolution speeds, and therefore longer integrations. In most of these integrations, we used the slowest plausible rate of tidal evolution of Namaka (about $7 \times 10^{-4} R_H~\rm{Myr}^{-1}$) that could produce its present eccentricity through smooth evolution within the 8:3 resonance over the Solar System's age. This rate was determined both from analytical considerations \citep{md99} and preliminary numerical runs, and they were in good agreement. Note that this rate assumes capture into the resonance just after formation (or even a primordial resonance) and smooth evolution into the present state. Despite being the slowest possible rate that still fits the constraints from the resonance, this tidal recession rate for Namaka is by a factor of few higher than our benchmark estimate of Namaka's tidal recession rate (Section 2).


We found that the capture efficiency was generally low ($\lesssim$ 20\%) and we could not identify the parameter combination that would lead to a much more robust capture probabilities. For capture to happen, Namaka's eccentricity had to be well below $0.01$, while the eccentricity of Hi'iaka could have a range of values, from very low to $0.01$, without affecting the probability of capture. We observe numerous sub-resonances of the 8:3 resonance (split due to precession induced by $J_2$ of Haumea), which are typically crossed in the direction from inclination-dominated ones (for example $8 \lambda_2 - 3 \lambda_1 -4 \Omega_1 - \varpi_1 $) to eccentricity dominated ones (like $8 \lambda_2 - 3 \lambda_1 - 5 \varpi_1 $). We find that the eccentricity-dominated arguments (at least in our setup) are stronger and lead to larger kicks (when crossed) and all the instances of capture observed in this work have been into eccentricity-dominated sub-resonances. 

\begin{figure}
\epsscale{.75}
\plotone{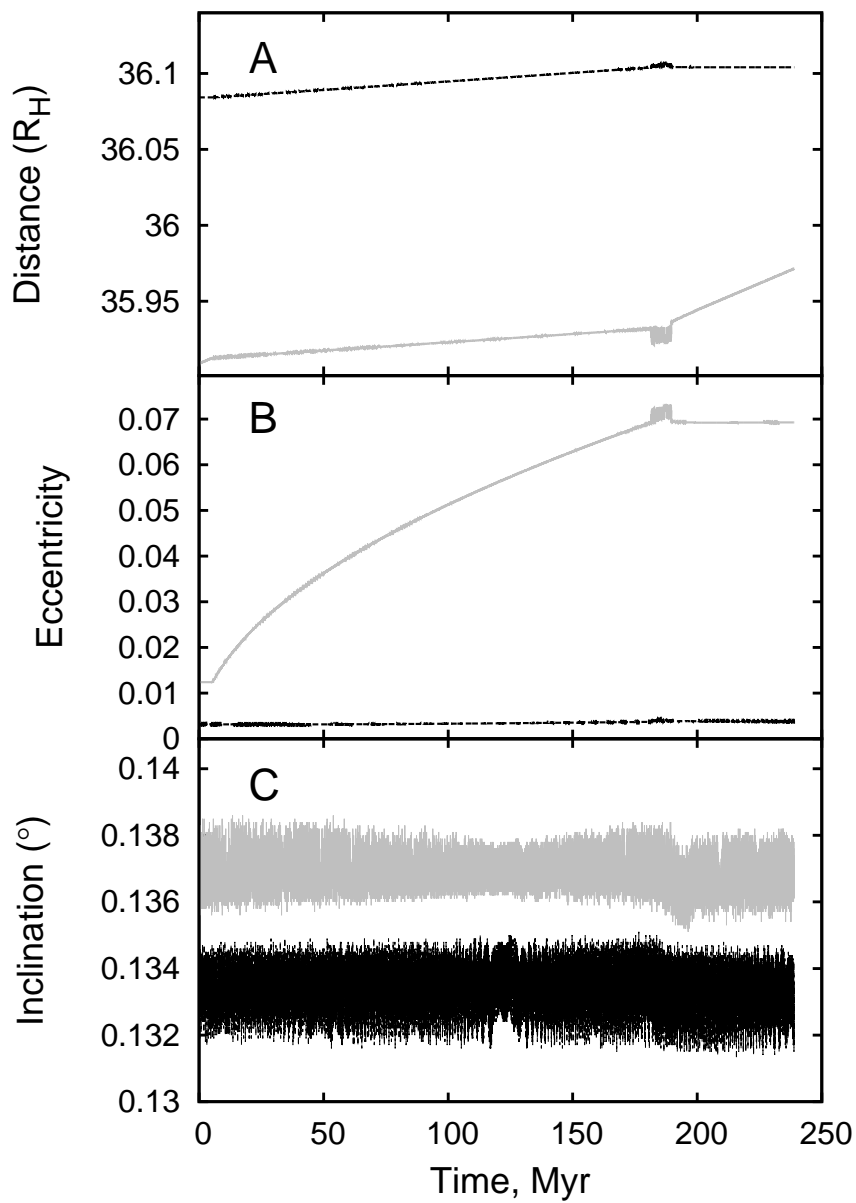}
\caption{One example of the evolution of Namaka (gray) and Hi'iaka (black) through their mutual 8:3 resonance, using nominal masses and dynamically "cold" orbits. Panel A plots semimajor axis of Namaka and the location of exact 8:3 commensurability with Hi'iaka, while panels B and C plot the moons' eccentricity and inclination relative to Haumea's equatorial plane. This simulation shows capture into a sub-resonance affecting only the eccentricity of Namaka, and the resonance breaks once the sub-resonances start overlapping.}
\label{83cap1}
\end{figure}

\begin{figure}
\epsscale{.80}
\plotone{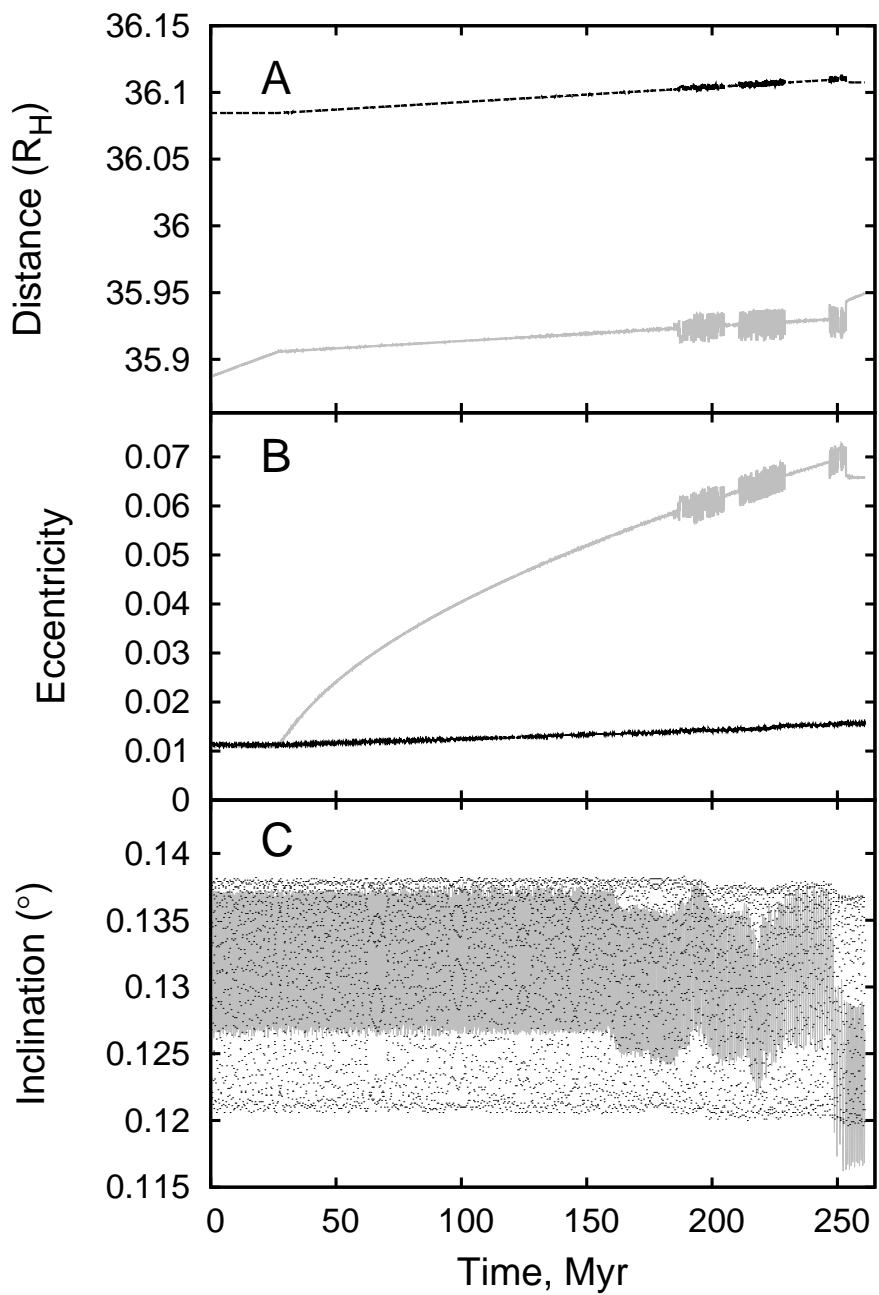}
\caption{Another example of the evolution of Namaka (gray) and Hi'iaka (black dots) through their mutual 8:3 resonance, using nominal masses and initially "cold" orbits. Panel A plots semimajor axis of Namaka and the location of exact 8:3 commensurability with Hi'iaka, while panels B and C plot the moons' eccentricity and inclination relative to Haumea's equatorial plane. This simulation shows capture into a sub-resonance affecting the eccentricities of both Namaka and Hi'iaka, and the resonance breaks once sub-resonances start overlapping.}
\label{83cap2}
\end{figure}

Fig. \ref{83cap1} shows the evolution of the moons' orbits in a case when capture was achieved. In this run, a pure Namaka eccentricity resonance is established (with the resonant argument $8 \lambda_2 - 3 \lambda_1 - 5 \varpi_1 $), followed by an increase in the eccentricity of Namaka but not Hi'iaka (with the inclinations not affected at all). Before the eccentricity could rise significantly, the sub-resonance overlap occurs and the moons' find themselves interacting through more than one resonant argument. This excites Hi'iaka's eccentricity but leads to decrease in Namaka's eccentricity. At this point, the system is only at the very edge of the chaotic zone, and these variations temporarily place Namaka's orbit just outside the original sub-resonance, effectively breaking the lock. Similar behavior is seen in Fig \ref{83cap2}, where the evolution through a combined Namaka-Hi'iaka eccentricity resonance ends once its sub-resonances start to overlap. 

A simple calculation using the approach of \citet{nes98} estimates the eccentricity at which the 5th order eccentricity-type resonance would overlap with the neighboring sub-resonance as:
\begin{equation}
e=\Bigl[ {J_2 \over 8} \Bigl({R_H \over a_1} \Bigr)^2 \Bigl( {3 M_H \over M_2}  \Bigl({a_2 \over a_1} \Bigr)^3 \Bigr)^{1/2} \Bigr]^{2/5}\approx 0.08
\end{equation}
which is close to the value of eccentricity at which we observe apparent sub-resonance overlap in Fig. \ref{83cap2}. Therefore we are confident that the sub-resonance overlap is the cause of the breaking of the resonance lock.

Simulations shown in Figs. \ref{83cap1} and \ref{83cap2} have been conducted using the nominal masses for satellites as reported by \citetalias{rag09}. In the previous section, we showed that the likely masses are lower by about a factor of two if the moons were to be in the stable 8:3 resonance for Gyr or so. However, in the previous section we assumed that there was no secular increase in the moons' eccentricities and inclinations, an assumption that is invalid during resonant evolution. So if a successful scenario of 8:3 resonant evolution were to be found, tidal evolution and stability would need to be studied simultaneously. However, given the results shown in Figs. \ref{83cap1} and \ref{83cap2}, we do not think that the moons could have obtained their eccentricities and inclinations solely by evolving through the 8:3 resonance, starting from "cold" orbits. We also conducted some simulations that assumed lower masses for the moons (in the range 50\%--90\% of the nominal ones), but we did not achieve capture. We estimate that the 8:3 capture efficiency drops below 10\% for masses that are significantly lower than nominal. Since lower-mass satellites would also have slower tidal evolution, analytically estimated rate of current tidal recession becomes an order of magnitude too short for what is required to excite both orbits to the observed state within 4 Gyr.   

We conclude that the capture of dynamically "cold" orbits into the 8:3 mean-motion resonance and subsequent tidal evolution within this resonance are unlikely to be the primary source of the moons' eccentricities and inclinations. The tidal evolution rates required are too high, capture resonance efficiencies are low, and, most importantly, the resonance is not continuously stable from "cold" orbits all the way to the present configuration. The first two problems are even more dire if the satellites' masses are lower than the \citetalias{rag09} nominal solution. Therefore it is necessary to explore other mechanisms for Haumea's moons to arrive into their mutual 8:3 MMR with the observed AMD.

\section{TNO-Hi'iaka Close Encounters and the 8:3 Mean-Motion Resonance}

In the previous sections we considered only tides raised by the moons on Haumea, with the resonances between the moons being the only possible sources of their orbital excitation. However, there is another well-established mechanism for perturbing orbits of TNO satellites: close encounters with other TNOs \citep{lev95, ste03}. Here we will use results of \citet{col08}, who explicitly addressed perturbations of passing TNOs on the orbit of Hi'iaka (fly-bys of Hi'iaka change the system's angular momentum much more than those involving Namaka). \citet{col08} find that the evolution of the TNO satellites' eccentricities is a form of random walk dominated by few major events known as "L\'evy flights". Due to the stochastic nature of this process, the range of plausible final eccentricities covers several orders of magnitude. \citet{col08} calculate that the most likely induced eccentricity of Hi'iaka is 0.005, an order of magnitude smaller than observed. However, their 66\% confidence interval includes eccentricities up to 0.03, and within 95\% confidence, final eccentricities for Hi'iaka can be as large as 0.2. These confidence limits arise from the stochastic nature of the process only, and do not take into account other possible sources of error (poorly constrained TNO size-distribution etc.). Therefore, generation of the observed angular momentum deficit through close encounters between Hi'iaka and passing TNOs, as previously suggested by \citet{sch09}, is a real possibility. 

\begin{figure}
\epsscale{.75}
\plotone{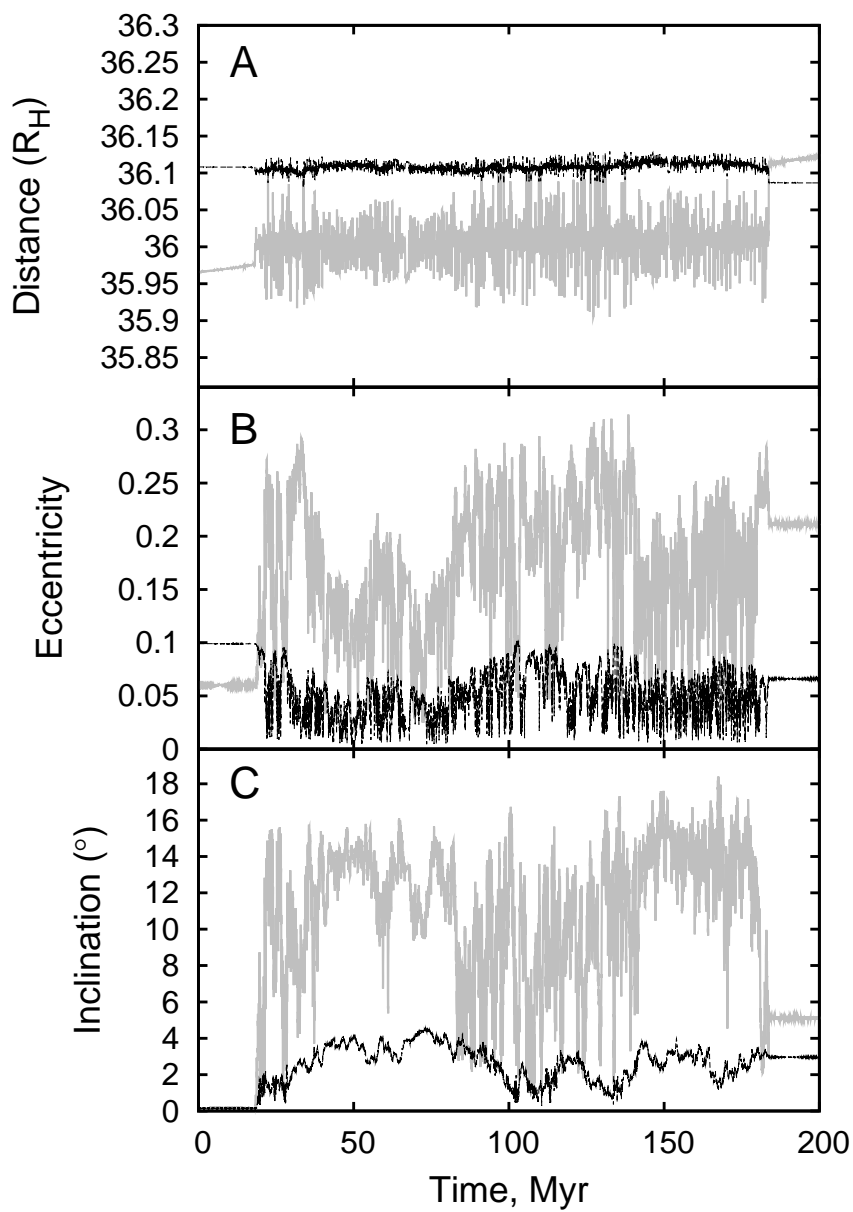}
\caption{An example of the evolution of Namaka into a 8:3 resonance with an initially excited Hi'iaka ($e_2=0.1$), using 50 \% of the nominal masses. Panel A plots semimajor axis of Namaka and the location of exact 8:3 commensurability with Hi'iaka, while panels B and C plot the moons' eccentricity and inclination relative to Haumea's equatorial plane. The moons leave the resonance in about 200 Myr, probably a  consequence of the Namaka's outward migration speed being a factor of few faster than the realistic rate.}
\label{hot}
\end{figure}

If Hi'iaka acquired its present (or comparable) eccentricity though close encounters, could it pass this orbital excitation to Namaka, and also bring the inclinations of both bodies to present values?\footnote{Accumulation of random impulse-like perturbations leads to inclinations and eccentricities in the approximate ratio $\sin{i}=e/2$. Given large uncertainties of the process, we will ignore any inclination of Hi'iaka produced by encounters, as it is unlikely to be as important as perturbations to eccentricity.} As we established in the previous Section, the 8:3 resonance is the best candidate for redistributing AMD between the two satellites. However, we still need to establish if a resonant encounter between a moderately eccentric Hi'iaka and a largely unexcited Namaka could lead to resonance capture and re-distribution of AMD between the moons. Figure \ref{hot} shows a numerical simulation where the moons were placed just wide of the 8:3 resonance and  Hi'iaka was given eccentricity of 0.1. Namaka subsequently tidally evolves into the resonance and capture into a chaotic resonance occurs, reminiscent of  long-term integrations in Section 3. Interestingly, the moons spend a short intervals in pure sub-resonances before returning to chaotic behavior spanning multiple sub-resonances. In any case, redistribution of moderate eccentricity of Hi'iaka to Namaka through the 8:3 resonance is fully consistent with the present eccentricities and inclinations of the moons. It is possible that this original eccentricity of Hi'iaka was generated by close encounters with TNOs, but also some amount of eccentricity leftover from formation cannot be excluded, in analogy with terrestrial planet formation \citep{cha98, agn99, obr06}. Last few collisions between bodies that accreted into Hi'iaka are likely to generate some eccentricity and inclination, and tidal or gas-drag damping are unlikely to have played a role during formation of Hi'iaka and Namaka.

In Fig \ref{hot} we assumed that the resonance was established through tides, after major TNO fly-bys of Hi'iaka. Of course, the fly-bys are an ongoing process, so it is possible for the resonance to be established or broken during these encounters. A major encounter which changes Hi'iaka's eccentricity by 0.01 is also likely to change its semimajor axis by about a percent, enough to break the resonance. This opens the possibility that the 8:3 resonance between the moons is not permanent but temporary and intermittent. It is also possible (but not required by available data) that the system is currently not in the 8:3 mean-motion resonance, but that the resonant exchange of AMD happened in the past. Our conclusions about the relatively low satellite masses required for the stability of the system depend on the resonance lasting for a long time (Gyr or so in the past), but do not require the resonance to be presently extant. Since the large-scale kicks to Hi'iaka's orbit from fly-bys should happen relatively infrequently, we will assume that the limits on satellite masses derived from long-term stability in the 8:3 resonance are applicable despite uncertainties about the history of the system. Similarly, it is not possible to say exactly how much of the present excitation of the system comes from encounters between TNOs and Hi'iaka and how much was added through subsequent evolution within the resonance. Long-term resonant evolution will invariably increase average eccentricities and inclinations, despite the chaotic nature of their short-term evolution. Given the results on TNO encounters from \citet{col08} and our own estimates of possible past tidal evolution of Namaka, we think that the dominant contribution to the moons' orbital excitation came from encounters and/or formation. 

To conclude, once both the 8:3 resonant interaction between the moons and the close encounters between Hi'iaka and external TNOs are taken into account, we are able to account for the present eccentricities and inclinations of Hi'iaka and Namaka. It is very likely the moons did form from a disk, rather than being direct collisional fragments, as \citet{sch09} proposed for Namaka. It is also clear that this proto-satellite disk would have to extend out to the present orbit of Hi'iaka, which would be unprecedented for collision-generated disks \citep{can04}. 

\section{Origin of the Satellites and the Haumea Family}

What are the implications of Haumea's moons' orbital history for the origin of the Haumea family? The sequence of events that led to the present Haumea system and the family is likely to be complex, but some conclusions can be made with confidence. The family-forming event must have happened in the present configuration of the outer Solar System, after the end of planetary migration \citep{lev08}. On the other hand, the collision in which Haumea itself formed and acquired its elongated shape and rapid rotation is very unlikely to have happened in the present Solar System. Only a slow collision between two similar-sized bodies appears to be able to reproduce Haumea's shape and the pure ice composition of the ejected fragments, which eventually  became satellites and/or family members \citep{lei10}. The type of slow collision explored by \citet{lei10} can also produce a low-dispersion-velocity family, but we must conclude that the extremely unlikely nature of this collision in the present Solar System rules out the possibility that the observed family had such an origin. Most likely, there must have been a number of early families, including Haumea's, that formed in the primordial, pre-migration trans-Neptunian space and that have been dispersed during planetary migration. Similarly, \citet{lei10} type event cannot form Hi'iaka and Namaka at their present locations. While \citet{lei10} find that numerous fragments orbit post-impact Haumea on eccentric orbits (commonly with $e=0.9$), these fragments do not carry sufficient angular momentum and will collisionally settle into a relatively compact disk close to Haumea. While it is possible that a wider exploration of parameter space may produce a larger disk, the results reported by \citet{lei10} are in agreement with other simulations of giant impacts which may lead to satellite formation.

The only way to bridge the gap between the end of the original collision (circum-Haumea material in a close orbit) and the family-forming event (some of the debris being ejected from the system with $\Delta v \simeq$ 150 m/s, with the rest forming a $70 R_H$-wide disk) appears to be the existence of at least one large past satellite which was formed in the original giant impact on Haumea and was subsequently collisionally destroyed \citep{sch09}. Unlike the larger collision that formed Haumea, collisional disruption of the satellite in the present Solar System is a reasonably likely event \citep{sch09}. The size of this ur-satellite can be estimated from the amount of material in Haumea's moons and family members, and \citet{sch09} estimate its diameter at 520~km, assuming that extant moons, family members and the ur-satellite all had the same density. For the density of $0.5 \rm{~g~cm^{-3}}$, this would imply a mass of $4 \times 10^{19}$~kg, or about 1\% of the mass of Haumea. Note that an icy  body of this size (comparable to Enceladus or Mimas) would be gravity dominated and have little porosity, so the actual parameters of the satellite were more likely to be $D=410$ km and $\rho=1 {\rm~g~cm^{-3}}$.

There are multiple constraints that the ur-satellite needs to satisfy in order to be a viable hypothesis. One important question is if the ur-satellite can evolve far enough from Haumea on reasonable timescales (Gyr or so) in order to serve as the progenitor of the current satellites. The other, related, issue is the amount of angular momentum that satellite could extract from Haumea without de-spinning it too much, as the dwarf planet is presently spinning relatively close to the break-up limit. 


When modeling the tidal evolution and the breakup of the ur-satellite, we should not ignore the possibility that its orbit was eccentric. \citet{lei10} found that bound fragments resulting from Haumea-making collision are typically on eccentric orbits, with $e \lesssim 0.9$. This initial eccentricity should dissipate as the fragments merge into a disk and accrete into one (or more) close-in satellites. Multiple satellite systems of this type are unstable, and smaller moons merging with (or scattering off) the ur-satellite could impart its orbit some initial eccentricity. During tidal evolution, this eccentricity could grow or damp, depending on tidal properties of Haumea and the ur-satellite. After \citet{can99}, we define parameter $A$ (subscripts $H$ and $s$ refer to Haumea and the satellite, respectively):

\begin{equation}
A={k_{2,s} Q_H \over k_{2,H} Q_s} {M_H^2 \over M_S^2} {R_s^5 \over R_H^5} 
\label{tidala}
\end{equation}

If $A < 19/28$ eccentricity can grow during tidal evolution. We can use $k_2 \sim \rho R^2$ \citep{md99} and $M \sim \rho R_{sphere}^3$, in which case
\begin{equation}
A={Q_H \over Q_s} {\rho_H \over \rho_s} {R_s \over R_{H, eq}} {R_{H, sphere}^6 \over R_{H, eq}^6}
\label{tidala2}
\end{equation}

Here we distinguish between the actual (equatorial) radius of Haumea that is applicable to the location of the tidal bulge, and the mean radius that determines volume. Assuming $\rho_s=1 {\rm~g~cm^{-3}}$ and $R_s=206$~km, we get that $A= 0.37~(Q_H / Q_s)$. Since both $Q$s are basically unknown (apart from likely being of the same order of magnitude), eccentricity could plausibly both grow and decay, with a slight tendency to excitation. In order to demonstrate that the ur-satellite hypothesis is tenable, we will assume $Q_H=50$, $Q_s=100$, and $Q_H/Q_s$=0.5, which is certainly plausible, if unconstrained by any available data. In this case $A=0.18$ and, according to Eq. 4 in \citet{can99}, ${\dot e} = 1.75 (e/a) {\dot a}$. We will use this estimate to determine the plausible eccentricity of the ur-satellite at different points of its orbital evolution.

How far out could have the ur-satellite evolved before it was destroyed? Integrating the tidal evolution equation \citep{md99}, we get

\begin{equation}
\Bigl({a \over R_H}\Bigr)=\Bigl({33 \over 2} {k_2 \over Q} {M_s \over M_H} \sqrt{GM_H \over R_H^3} \tau \Bigr)^{2/13}
\label{ursat}
\end{equation}

If we use $k_2=0.5$ and $Q=50$ for Haumea, a mass ratio of $0.01$ and a time $\tau=2$~Gyr, we get $a=34~R_H$. However, if the eccentricity of the satellite in the late stages of this evolution was large, tidal evolution would have been significantly faster as stronger interactions at pericenter dominate over weaker ones at apocenter. The eccentricity of $0.5$ produces more than an order of magnitude faster tidal evolution than that of a circular orbit of the same semimajor axis \citep[by a factor of $(1+15/2 e^2)/(1-e^2)^6$, to order $e^2$; ][]{mac64}. This acceleration would lead to a 50\% larger semimajor axis reached during the same time, i.e. $a=50 R_H$ over 2~Gyr. While the ur-satellite's eccentricity may have been lower when close to Haumea, the overwhelming majority of tidal evolution is always spent during slow evolution at large distances. Using our estimate of eccentricity growth, we find that semimajor axis expansion from 5 to 50~$R_H$ would also lead to $\simeq50$-fold growth of eccentricity, so the ur-satellite final eccentricity of 0.5 is actually consistent with low post-formation eccentricity of only 0.01.

\begin{table}

\begin{center}
\caption{Properties of Haumea and high-resolution simulations of impacts on Haumea by \citet{lei10}. The columns are mass, the three principal radii, spin period and the relative angular momentum (in terms of critical momentum $L_C$).\label{zoe}}
\bigskip
\begin{tabular}{crrrrrr}
\tableline\tableline
Example & M [$10^{21}$~kg] & $R_a$ [km] & $R_b$ [km] & $R_c$ [km] & $P_{spin}$ [hr] & $L/L_c$\\
\tableline
Run 1 &	 4.3 &	960 &	870 &	640 &	3.6 & 0.839 \\	
Run 2 & 4.3 &	1090 &	820 &	680 &	3.4 & 0.931 \\
Run 3 & 4.3 &	1100 &	940 &	640 &	3.7 & 0.976 \\
Run 4 & 4.2 &	1100 &	810 &	620 &	3.9 & 0.832\\
Average & & & & & & 0.894\\
Haumea & 4 &	980 &   759 &	500 &	3.92 & 0.756 \\
\tableline
\end{tabular}
\end{center}
\end{table}

The ur-satellite could therefore reach a a semimajor axis as large as $50 R_H$, and with an eccentricity of 0.5 it would spend much of its time beyond $70~R_H$, where Hi'iaka's orbit is now. Is this consistent with the present rapid rotation of Haumea? There are two separate  questions here: how fast could Haumea possibly have rotated (without flying apart), and what would be its likely spin rate after formation. As the spin rate is correlated with the (spin-dependent) rotational deformation of the object, it may be better to use angular momentum as a basic quantity for such comparisons. The estimate of the maximum possible angular momentum by \citet{sch09} as being twice the present is not applicable, as they directly compare the spin rate of the current triaxial Haumea to the theoretical breakup spin rate of an idealized spherical body of  the same mass and density. Using a Jacobi ellipsoid as a model is also not useful, as a Jacobi ellipsoid has uniform density and can in principle have infinite angular momentum, which is not applicable to real bodies. More generally, treating Haumea as a fluid body is probably not justified \citep{hol07}. For our purposes, it may be more productive to try to answer the second question, about the likely post-formation spin. 

The best estimate of Haumea's initial angular momentum can be obtained from high resolution formation simulations by \citet{lei10}. While their final planets have spins very close to the present spin of Haumea (by design), they also have lower densities, so they are actually closer to breakup than the real Haumea. Here we will use $L_c=C_H M_H R_{sphere}^2 \omega_c$ as the unit of angular momentum, where dimensionless moment $C_H=0.33-0.4$ quantifies central concentration of mass in Haumea (we assume $C_H$ is constant with rotational deformation), $M_H$ is Haumea's mass, $R_{sphere}=\sqrt[3]{(3 M_H)/(4\pi \rho_H)}$ is its mean radius, and $\omega_c=\sqrt{(4\pi G \rho_H)/3}$ is the critical spin rate of spherical Haumea ($\rho_H$ is its density, also assumed to be constant with spin). Table \ref{zoe} shows the ratio between the actual and unit angular momentum for Haumea and four high-resolution simulations of Haumea from \citet{lei10}. The simulations have scaled angular momentum that is 83-97\% of the critical limit, while Haumea's is only $0.76 L_c$, so Haumea's original angular momentum could have been 10-25\% larger. Assuming $C_H=0.35$ and $M_s=0.01 M_H$, initial $a_s= 5~R_H$, $e_s=0.01$ and final $a_s=50R_H$ and $e_s=0.5$, the angular momentum needed for this tidal evolution is about $0.11 L_c$. It is therefore entirely plausible that some of Haumea's original rotational angular momentum was transfered to the ur-satellite. In other words, Haumea today appears to be less close to breakup than the formation simulations by \citet{lei10} suggest, implying that some amount of spindown happened since formation, and tidal evolution of a large past satellite is the prime suspect for this despinning. Additionally, recent work by \citet{cuk12} and \citet{can12} shows that formation of a sizable satellite is compatible with a fast-spinning primary, at least in case of the Earth-Moon system.

If the ur-satellite was destroyed by a collision with another, smaller, TNO, what do we expect the distribution of fragments' velocities to be? If the impactor has about $10^{-3}$ of the ur-satellite's mass and it impacted at 3~km/s \citep{sch09}, it could have imparted only a few meters per second to the center of mass of the debris field, so the impactor's contribution to total angular momentum can be ignored. Dispersion from the breakup would add considerable velocity to the fragments, on the order of the ur-satellite's escape velocity of 160 m/s. With the escape velocity (from the Haumea system) being only about 100 m/s at 70 $R_H$, a large fraction of fragments would certainly escape, in agreement with the existence of the Haumea family. Depending on the position of the ur-satellite in its orbit, it would have an orbital velocity of about 50-100 m/s, with the slowest motion taking place when it was at apocenter. It is clear from these numbers that the orbital motion would introduce a significant asymmetry of the fragment velocities (with respect to Haumea). The fragments launched in the direction of ur-satellite's orbital motion would easily escape the system with $V \simeq 200$ m/s, while those launched opposite the orbital motion would have a good chance of staying bound, on retrograde orbits. On the other hand, any very slow-moving fragments would be retained on prograde orbits similar to that of the ur-satellite.

\citet{rag07}, \citet{lyk12} and \citet{vol12} found that most of the family's mass is contained in several large members, and that they all (except 1996 TO$_{66}$) have minimum ejection velocities of about 100 m/s or so. Ejection from a distance of $70 R_H$ implies an even larger velocity from the ur-satellite's center of mass, before escaping Haumea's gravitational pull. The fragment that we think was progenitor of Hi'iaka and Namaka (carrying about a quarter of total mass) would be required to have a velocity less than 50 m/s to stay bound to Haumea on a prograde orbit. On the other hand, if it was launched approximately opposite the ur-satellite's orbital motion, it would end up on a bound retrograde orbit after having the ejection velocity typical of the family. 

While the size-distribution of the family indicates that the satellites are most likely descendants of a single large fragment, they cannot be direct fragments from the breakup \citep[as suggested by ][]{sch09}. Once the likely resonant interaction is taken into account, their initial orbits were almost certainly too circular and coplanar for primary fragments. Most likely, some of the smaller bound debris (probably on highly eccentric and inclined orbits) collided with the largest fragment, setting off a collisional cascade which ended with the formation of a flat disk in the plane of Haumea's equator. Hi'iaka and Namaka then formed from that disk on circular and coplanar orbits that could have been either prograde or retrograde relative to Haumea's spin, depending on the ejection circumstances of the largest retained fragment. Since we do not know the exact distribution of fragments' velocities, we cannot say if prograde or retrograde disk is a more likely outcome. While very small total ejection velocities are less likely (if the distribution was something like Maxwellian), they always result in retention. Velocities required to produce retrograde bound fragment are more likely but only a fraction of fragments (i.e. those launched roughly against the ur-satellite's orbital motion) would be retained.

We conclude that if Hi'iaka and Namaka originated in the disruption of a past satellite, their present orbits may plausibly be retrograde relative to Haumea's spin. While this may seem like a radical hypothesis, there is nothing about the observations of Haumea and the moons to distinguish between prograde and retrograde orbits. At present, only the direction of motion of the moons is known, while Haumea is unresolved and its rotation is known from lightcurve alone. The perturbations on satellite orbits suggest a rotational axis that is close to being perpendicular to the orbit of Hi'iaka, without any indication of the direction of its rotation. It is indicative that our in-depth exploration of the moons' orbital evolution (Sections 2-5) could not find any evidence of convergent orbital evolution. The eccentricity-generating mechanism that we find most consistent with the moons' orbits are TNO flybys of Hi'iaka, which also cause a random walk in Hi'iaka's semimajor axis. These flybys are insensitive to the orbit of Hi'iaka being retrograde or prograde. Subsequent entry into the chaotic 8:3 resonance which likely redistributed angular momentum deficit between the moons could happen due to Hi'iaka's random walk in $a$, or by Namaka's tidal evolution {\it in either direction} (see the following Section). 

While it will be challenging to determine observationally the alignment or anti-alignment of Haumea's rotation and the moons' orbital motion, such an observation is plausible. One promising approach would be to detect Rossiter-McLaughlin effect \citep{oht05, win05} during transits of Namaka in front of Haumea, which may be observable using most advanced ground-based instruments. Apart from the Rossiter-McLaughlin effect, three additional methods for determining the sense of Haumea's rotation are worth mentioning. It is known that there is a somewhat localized dark red spot on Haumea \citep{lac08}. If this spot is constrained to a small region of Haumea's surface, then its motion can be used to track the spin direction of Haumea. Its location and motion can be determined by 1) detecting a localized color-anomaly during a mutual event of Namaka or 2) measuring the astrometric center-of-light wobble due to the rotation into and out of view of the red spot. The size of the wobble depends on the color and localization of the red spot, which are degenerate in the current lightcurve solutions \citep{lac08}. HST observations in a well-chosen filter of a mutual event and the requisite out-of-event data are the best method for detecting these effects and may be precise enough to determine whether Haumea's rotation is prograde or retrograde with respect to the satellite orbits. The third method is to directly detect the effect of the triaxial shape of Haumea in the satellite's orbits, i.e., measurement of the $C_{22}$ of Haumea. \citet{rag09} found this to be very difficult, but further investigation may be warranted. We urge the observers to attempt such observations, as a confirmation of retrograde orbits would put strong constraints on the origin of Hi'iaka and Namaka, as well as Haumea's heliocentric family.


\section{Dynamics of Retrograde Hi'iaka and Namaka}

In this Section we will briefly discuss some implications of Hi'iaka and Namaka having retrograde orbits.

The only substantial effect on their dynamics would be the opposite direction of their tidal evolution. While prograde Namaka and Hi'iaka evolve away from Haumea while converging, retrograde Hi'iaka and Namaka must involve inward, and their orbits would be diverging in the process. Given enough time, both moons should eventually spiral onto Haumea if they are retrograde. Our estimates of plausible amount of tidal evolution (Section 2) still apply, and it is unlikely that Namaka moved inward by more than a few percent of its distance, and Hi'iaka not even that much. Note that inward tidal evolution (unlike outward one) is a runaway process that constantly accelerates, so the very existence of an extant retrograde (or super-synchronous) satellite suggests that its orbit was unlikely to have evolved much in the past, unless we are lucky to observe it soon before its final plunge (as is the case with Phobos). All of the above considerations would suggest that tidal evolution played only minor role in the histories of Hi'iaka and Namaka.

Divergent rather than convergent migration has profound implications for resonance crossings, as convergent orbits usually get captured in MMRs, while divergent ones do not. There are two separate questions about the potential divergent migration of Haumea's moons: can they still enter the 8:3 (which they may presently occupy), and can their orbits be excited by divergent resonance crossings? 
\begin{figure}
\epsscale{.75}
\plotone{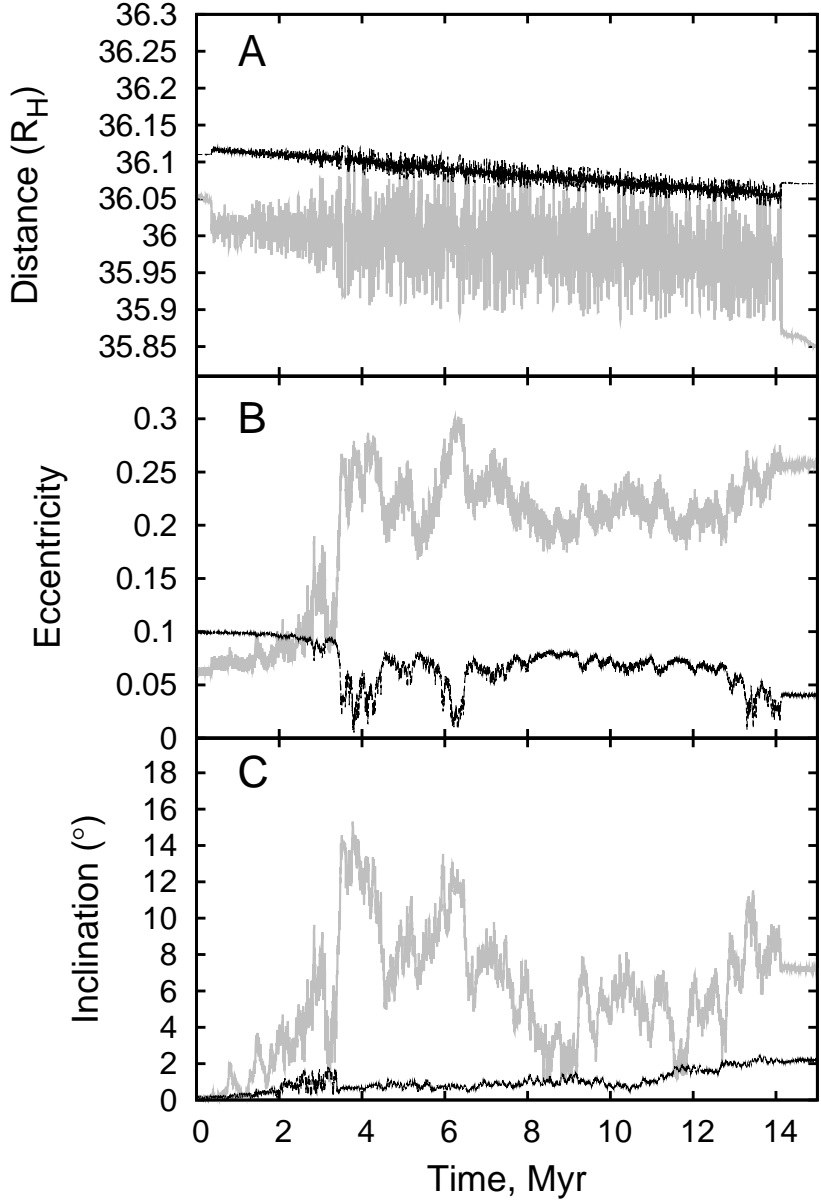}
\caption{Inward tidal evolution of (retrograde) Namaka into a 8:3 resonance with an initially excited Hi'iaka ($e_2=0.1$), using 50 \% of the nominal masses. Panel A plots semimajor axis of Namaka and the location of exact 8:3 commensurability with Hi'iaka, while panels B and C plot the moons' eccentricity and inclination relative to Haumea's equatorial plane. The moons leave the resonance in about 15 Myr, as Namaka's evolution rate is two orders of magnitude faster than the realistic rate. Both during and immediately after the resonance, the moons' orbits in the simulation are consistent with the observed system.}
\label{reverse}
\end{figure}

In Section 5 we discussed how 8:3 resonance can redistribute AMD between Hi'iaka and Namaka, if the pre-resonance orbit of Hi'iaka had a moderate eccentricity (presumably generated in close encounters with other TNOs). Fig \ref{hot} shows how the orbits can enter 8:3 resonance through convergent migration, with eccentricities and inclinations evolving toward their present values. Would this work in the case of divergent migration? Figure \ref{reverse} shows the inward migration of Namaka into the mutual 8:3 resonance. Once the resonance is reached, both orbits become chaotic and the familiar pattern of eccentricity and inclination variation is established, with the present orbital elements being fairly representative of the range within which eccentricities and inclinations vary. Alternatively, random walk in the orbit of Hi'iaka due to encounters may also place the moons' orbits directly in the 8:3 resonance, and lead to the present distribution of orbital elements. In any case, present (or at least long-lived) residence in the 8:3 resonance appears to be as likely with retrograde Hi'iaka and Namaka as it is in the prograde case.

\begin{figure}
\epsscale{.75}
\plotone{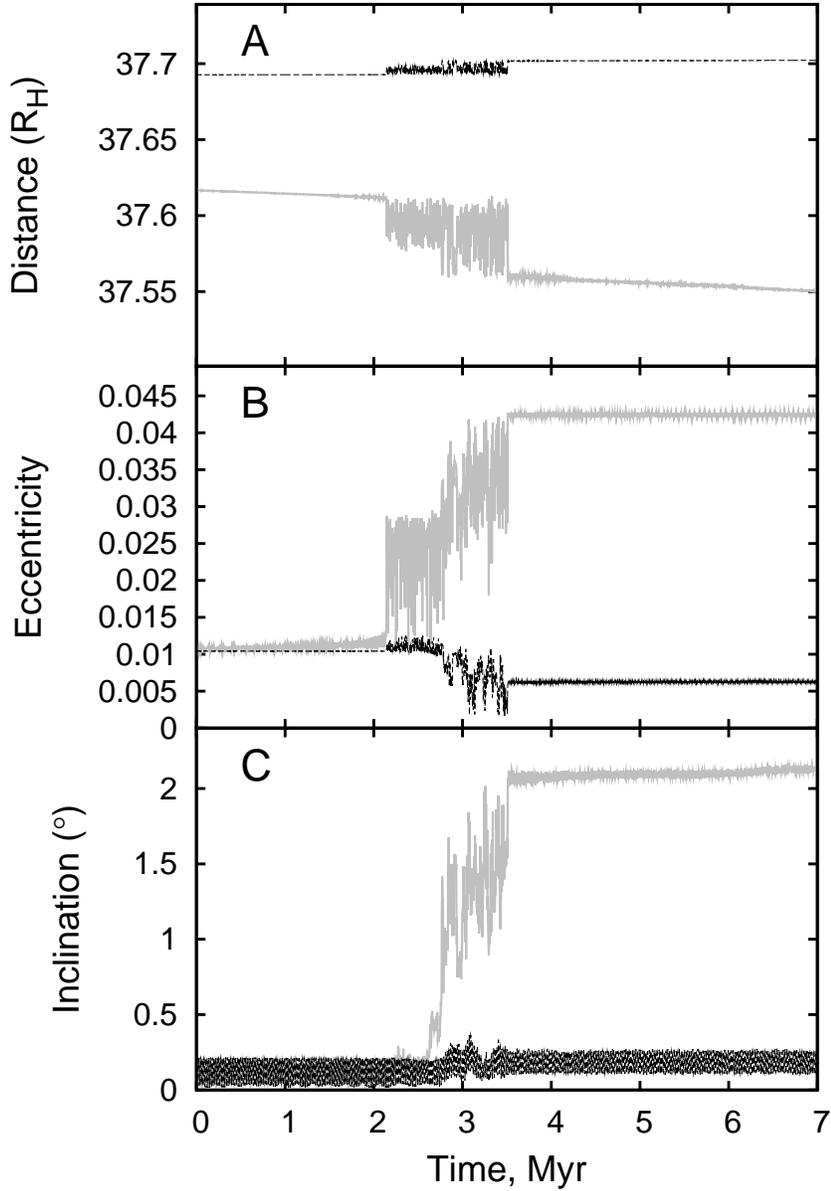}
\caption{Inward tidal evolution of (retrograde) Namaka through the 5:2 resonance with an initially low-eccentricity Hi'iaka ($e_2=0.01$), using 50 \% of the nominal masses. Panel A plots semimajor axis of Namaka and the location of exact 5:2 commensurability with Hi'iaka, while panels B and C plot the moons' eccentricity and inclination relative to Haumea's equatorial plane. As the moons cross the resonance, Namaka's acquires eccentricity and inclination that are well short of the observed values. Namaka's semimajor axis evolution rate is about one order of magnitude faster than the realistic rate.}
\label{52evo}
\end{figure}
When divergent orbits encounter mean-motion resonances, the orbits typically jump over the resonance, and their eccentricity and/or inclinations experience a "kick" \citep{chi02}. There are some exceptions for migration of already excited orbits into chaotic resonances (like the case shown in Fig. \ref{reverse}), but this rules almost always holds for low-eccentricity, planar orbits encountering major mean-motion resonances. Would it be possible that Hi'iaka and Namaka's orbits were excited through a divergent resonance passage, removing a need for TNO encounters? The best candidate would be their mutual 5:2 resonance, which is outside the 8:3 resonance and may be within the plausible range of Namaka's past tidal evolution. In Figure \ref{52evo} we show a passage of Namaka through the mutual 5:2 mean-motion resonance, starting with low $e$ and $i$ orbits. Only moderate excitation of Namaka's orbit is observed, and Hi'iaka's eccentricity even drops somewhat (resulting angular momentum deficit is two orders of magnitude below the observed one). This simulation was somewhat accelerated relative to the expected realistic rate, but comparison with simulations using a variety of orbit divergence speeds indicates that this result is not dependent on the evolution rate.

We also explored the effect of the 8:3 resonance on dynamically cold orbits and found that there was barely any observable excitation. Stronger 2:1 and 3:1 resonances may provide more excitation, but the former is likely beyond reach of Namaka's tidal evolution, and the other one is inward of the present position of Namaka. In both cases, well-placed encounters between a moon (likely Hi'iaka) and a passing TNO could change the period ratio by more than the total of Namaka's tidal evolution, but such encounters would also generate orbital excitation all by themselves, without a need to invoke resonances.

An interesting consequence of the satellites being retrograde is a possibility that there were more moons interior to Namaka that have since collapsed onto Haumea due to tidal forces. A satellite the size of Namaka would have spiraled into Haumea if it started at a distance smaller than about 25 $R_H$, and a ten times smaller satellite would not survive interior to 18 $R_H$. Resonances between these putative lost moons and Namaka would be an alternative source of orbital excitation, but we we will not explore this possibility further here as there is no direct evidence that these additional moons ever existed.

In conclusion, regardless of the moons being prograde or retrograde, the most likely sources of their orbital excitation were encounters between Hi'iaka and passing TNOs, or Hi'iaka's possible late-stage accretion of additional moonlets. Subsequently, angular momentum deficit was passed to Namaka through the chaotic 8:3 resonance, which may still be active.

\section{Conclusions}

We have studied the long-term dynamics of the two satellites of the dwarf planet Haumea using numerical integrations, and we have reached the following conclusions:

1. The eccentricities and inclinations of Hi'iaka and Namaka strongly suggest that the two moons at some point participated in a mean-motion resonance of order 2 or higher, and the observations are consistent with the moons currently being in the chaotic 8:3 mean-motion resonance. This process is independent from the mechanism of the initial orbital excitation, as high-order resonances can re-distribute pre-existing excitation between the moons' eccentricities and inclinations.

2. The eccentricities and inclinations are unlikely to have been generated by smooth convergent migration into mean-motion resonance, starting with close-in, circular and planar orbits. This applies both to 3:1 and 8:3 resonance. Analytical estimates of past tidal evolution over the age of the Solar System imply that the satellites have experienced only limited orbital change compared to their present orbits. This suggests that the satellites did not achieve their present positions from close-in orbits that would be expected from a typical post-collision debris disk.

3. If the moons spent hundreds of Myr in the 8:3 resonance with present eccentricities and inclinations, their densities and masses are likely only half of those reported by \citetalias{rag09}. Larger masses invariably lead to instability and scattering on 100-Myr timescales.

4. The best supported orbital history suggests Hi'iaka's eccentricity was excited by encounters with other TNOs \citep[][]{col08, sch09}, with Namaka's orbit subsequently excited through 8:3 resonance with Hi'iaka. The resonance would have been encountered through either limited tidal evolution of Namaka or encounter-driven random walk of Hi'iaka.

5. Hi'iaka and Namaka appear to have formed close to their present distances from Haumea, on low-eccentricity, planar orbits.

6. We find that the currently best explanation for the origin of Haumea's satellites and compositional family is a collisional disruption of a past large moon of Haumea \citep{sch09}. We estimate that this "ur-satellite" could have had a mass of $0.01 M_H$, semimajor axis of $\simeq 50 R_H$ and an eccentric orbit $e=0.5$. This scenario is consistent with Haumea's overall angular momentum budget if it formed in a large, slow collision \citep{lei10}. The escaping debris formed the Haumea family, while the debris bound to Haumea formed a large disk from which Hi'iaka and Namaka formed. 

7. We find that the bound material from such a disruption is sometimes retrograde, meaning that the present satellites may orbit opposite the spin of Haumea. While surprising, this result would be consistent with both the observations and our own long-term dynamical simulations. This possibility of retrograde orbital direction of Hi'iaka and Namaka offers a direct way of confirming the disrupted satellite hypothesis. Prograde rotation would not falsify the disrupted satellite scenario but would also be consistent with the previously proposed giant impact hypothesis for the origin of the system \citepalias{rag09}.

\acknowledgments

M. {\'C} is supported by NASA's Outer Planets Research Program award NNX11AM48G. We wish to thank an anonymous reviewer for providing us with useful comments and suggestions.


\nopagebreak

\end{document}